\documentclass[twocolumn,superscriptaddress]{revtex4-2}

\usepackage{chemformula} 
\usepackage[T1]{fontenc} 
\usepackage{graphicx}
\usepackage{bm}
\usepackage{subfigure}
\usepackage{float}
\usepackage{braket}
\usepackage{xcolor}
\usepackage{hyperref}
\usepackage{amsmath}
\usepackage{yhmath}
\usepackage{xspace}

\newcommand{\psh}[2]{\ensuremath{\langle #1|#2\rangle}\xspace}
\newcommand{\bmtheta}{{\bm \theta}}
\newcommand{\bfw}{{\bf w}}

\def\ddroit{{\rm d}}

\newcommand{\bfr}{\mathbf{r}}

\newcommand{\bfv}{\mathbf{v}}

\begin{document}

\title{How an Equi-ensemble Description Systematically Outperforms the Weighted-ensemble Variational Quantum Eigensolver}

\author{Akilan Rajamani}
\affiliation{ICGM, Université de Montpellier, CNRS, ENSCM, Montpellier, France.}

\author{Martin Beseda}
\affiliation{Dipartimento di Ingegneria e Scienze dell’Informazione e Matematica, Università dell’Aquila, Via Vetoio, I-67010 Coppito-L’Aquila, Italy.}

\author{Benjamin Lasorne}
\affiliation{ICGM, Université de Montpellier, CNRS, ENSCM, Montpellier, France.}

\author{Bruno Senjean}
\email{bruno.senjean@umontpellier.fr}
\affiliation{ICGM, Université de Montpellier, CNRS, ENSCM, Montpellier, France.}%

\begin{abstract}
    Calculating excited states in chemistry is crucial to provide insight into photoinduced molecular behavior beyond the ground state, enabling innovations in spectroscopy, material sciences, and drug design.
While several approaches have been developed to
compute excited-state properties,
finding the best ratio between computational cost and accuracy remains challenging.
The advent of quantum computers brings new perspectives, with the development of quantum algorithms that promise an advantage over classical ones.
Most of these new algorithms are inspired from
previous classical ones, but with different
pros and cons.
In this Letter, we focus on the
generalization of
the variational principle for many-body excited-states that led to the ensemble variational quantum eigensolver (VQE).
We compare the performance of two ensemble VQE approaches, the equi-ensemble and weighted-ensemble ones, and conclude that the equi-ensemble is the way to go.
\end{abstract}

\maketitle

Let us consider an ensemble of $K$ initial orthonormal states $\lbrace \ket{\Phi_j} \rbrace_{j=0,\hdots,K-1}$ transformed by the same variationally parametrized unitary operator $\hat{U}(\bmtheta)$.
We define the ($\bfw$-weighted) ensemble energy as~\cite{theophilou1979energy,gross1988rayleigh}
\begin{eqnarray}
E_C^{\bfw}(\bmtheta) &=& \sum_{j=0}^{K-1} w_j \bra{\Phi_j} \hat{U}^\dagger(\bmtheta) \hat{H} \hat{U}(\bmtheta) \ket{\Phi_j} \nonumber \\
&& \geq \sum_{j=0}^{K-1} w_j E_j,
\label{eq:cost}
\end{eqnarray}
where $w_j \geq w_{j+1}$, $\hat{H}$ is the electronic structure Hamiltonian, and $E_j \leq E_{j+1}$ its exact eigenvalues. The inequality in Eq.~(\ref{eq:cost}) reflects the
variational principle generalized to an ensemble of states, such that minimizing
the ensemble energy with respect to the parameters $\bmtheta$ will lead to
$\hat{U}(\bmtheta) \ket{\Phi_j} = \ket{\Psi_j(\bmtheta)} \sim \ket{\Psi_j}$, $\ket{\Psi_j}$ being the $j$th eigenvector of $\hat{H}$ with eigenvalue $E_j$,
assuming that the unitary $\hat{U}(\bmtheta)$ can achieve this goal.

This generalized variational principle has been
used to extract excited-state properties from 
ensemble density functional theory ~\cite{gross1988density,
filatov1999spin,kazaryan2008excitation,filatov2015spin,
filatov2015ensemble,filatov2017analytical,
liu2021analytical,filatov2021description,
gould2017hartree,gould2018charge,
gould2019density,gould2020density,
gould2020approximately,gould2020ensemble,
gould2021ensemble,gould2021double,sagredo2018can,
senjean2015linear,senjean2016combining,
deur2018exploring,deur2019ground,
senjean2018unified,marut2020weight,
cernatic2022ensemble,cernatic2024extended,
scott2024exact,
cernatic2024neutral,dupuy2025exact,
fromager2025ensemble,gould2025variational,
gould2025ensemblization,daas2025ensemble}
and more recently in the context of the one-particle reduced density-matrix functional theory~\cite{schilling2021ensemble,
liebert2021foundation,
liebert2023deriving,
liebert2023exact},
variational Monte--Carlo~\cite{schautz2004optimized,
wheeler2024ensemble} and
density-matrix embedding methods~\cite{cernatic2024fragment}.
Turning to multiconfigurational approaches,
the generalized variational principle is used in conjunction with configuration interaction (CI) and
multiconfigurational self-consistent field (MCSCF) approaches~\cite{helgaker2014molecular}.
For the latter, the variational principle is used not only for the linear many-body variational problem (CI/MC) but also for the nonlinear one-body variational orbital-optimization problem (SCF). In general, 
the equi-ensemble ({\it i.e.}, $w_j = w_{j+1}$, $\forall j$) version of Eq.~(\ref{eq:cost}) is
considered for the SCF orbital optimization to provide
a democratic description of all the MC states of the ensemble, which is especially important when vibronic couplings are large, {\it i.e.}, in the vicinity of avoided crossings and conical intersections~\cite{domcke2004conical}, thus leading to the so-called state-average MCSCF (SA-MCSCF).
Recently, the quantum computing analog of the SA-MCSCF method has been derived and called state-average orbital-optimized variational quantum eigensolver (SA-OO-VQE)~\cite{yalouz2021state,
yalouz2022analytical,
beseda2024state,
illesova2025transformation}, inspired by the previous work of Nakanishi and coworkers on the subspsace-search VQE~\cite{nakanishi2019subspace}, which is equivalent but without orbital optimization.
SA-OO-VQE is analogous to SA-MCSCF as regards the state-average orbital optimization problem, which is performed classically. However,
while SA-MCSCF always uses an actual -- full or partial, but supposedly exact (numerically) for the targeted ensemble -- diagonalization algorithm for the many-body part ({\it i.e.}, for $\hat{U}(\bmtheta)$), SA-OO-VQE uses an
approximate parameterized-circuit ansatz $\hat{U}(\bmtheta)$ to minimize
the ensemble energy.
How this ansatz --
for which several different expressions have been derived~\cite{Peruzzo2014,
yungTransistorTrappedionComputers2014,
barkoutsosQuantumAlgorithmsElectronic2018,
romeroStrategiesQuantumComputing2018,
leeGeneralizedUnitaryCoupled2019,
evangelistaExactParameterizationFermionic2019,
sokolovQuantumOrbitalOptimizedUnitary2020,
khamoshiAGPbasedUnitaryCoupled2022,
smartQuantumSolverContracted2021,
marecatRecursiveRelationsQuantum2023,
materiaQuantumInformationDriven2024,
kandala2017hardware,
ganzhornGateEfficientSimulationMolecular2019,
ogormanGeneralizedSwapNetworks2019,
burtonAccurateGateefficientQuantum2024,
gardEfficientSymmetrypreservingState2020,
sekiSymmetryadaptedVariationalQuantum2020,
anselmettiLocalExpressiveQuantumnumberpreserving2021,
lacroixSymmetryBreakingSymmetry2023,
ruizguzmanRestoringSymmetriesQuantum2024} --
can actually achieve diagonalization without any convergence issues is one
of the biggest challenges of VQE~\cite{mccleanBarrenPlateausQuantum2018,
bittelTrainingVariationalQuantum2021,
dcunhaChallengesUseQuantum2023,
Larocca_2025,
leonePracticalUsefulnessHardware2024}.
Such a difference is also not without consequences 
as regards the choice of the initial orthonormal states $\lbrace \ket{\Phi_j} \rbrace_{j=0,\hdots,K-1}$,
and
has led to a revival of the use of ensemble theory to extract many-body excited states in the context of quantum computing~\cite{nakanishi2019subspace,
parrish2019quantum,
yalouz2021state,
yalouz2022analytical,
omiya2022analytical,
bierman2022quantum,
benavides2022excitations,
xu2023concurrent,
heya2023subspace,
beseda2024state,
chen2024crossing,
bierman2024qubit,
benavides2024quantum,
hong2024refining,
ding2024ground,
illesova2025transformation}.

Let us now discuss in more details the role of the weights $\bfw$ in Eq.~(\ref{eq:cost}).
To converge towards the eigenvectors of the problem,
while the weights $\bfw$ do not need to take particular values in principle 
apart from being sorted in descending order ($w_j > w_{j+1}$),
they can have a huge impact in practice.
Indeed, the choice of the weights can strongly influence the parameter landscape and impact the
minimization of the cost function $E_C^\bfw(\bmtheta)$, as shown recently
in Refs.~\citenum{hong2024refining} and \citenum{ding2024ground}.
Interestingly, they also have an impact in ensemble DFT when approximate Hartree-exchange-correlation functionals of the density are used~\cite{senjean2015linear}, such that they are sometimes taken as Boltzmann weights (with a fictitious temperature arbitrarily fixed by the user), thus leading to thermal DFT~\cite{PRL11_Pittalis_exact_conds_thermalDFT,
PRB16_Pribram-Jones_AC_thermalDFT,
pastorczak2013calculation}. Note that
Boltzmann weights may also be used in SA-MCSCF to avoid active-space discontinuities with respect to a finite adiabatic target subspace at various geometries~\cite{olsen2015canonical}.


Along the lines of the aforementioned works,
we question the use of different weight values
in practical calculations.
Using descending weights requires a decision to be made in advance as regards the guess (initial states and associated weights) to be transported to the final eigenstates which are not known in advance.
If the guess does not 
match the final eigenenergy order, which cannot be predicted before being calculated,
there is a source of problem
that we address in this Letter.
Indeed, we will show in the following that -- at least in the context of quantum chemistry on quantum computers studied herein --
the additional complexity arising from considering ordered weights is highly detrimental
and generally prevents convergence towards the
targeted subspace spanned by the $K$ lowest eigenstates.
Such a failure makes the final results unreliable.
Some attempts have been proposed to mitigate the convergence issue, for instance, by relaxing the use of different $\hat{U}(\bmtheta)$ per circuit and imposing the
orthonormality constraint explicitly 
in the cost function rather than implicitly in the unitary ansatz, at the expense of performing Hadamard tests for each cost function evaluation~\cite{bierman2022quantum}.

Alternatively, a particular limit of the ensemble variational principle in Eq.~(\ref{eq:cost})
is the equi-ensemble one, when $w_j = w_{j+1}$, $\forall j$~\cite{fan1949theorem}.
In such a case, the minimization of Eq.~(\ref{eq:cost}) does not converge toward the individual
$K$ lowest eigenstates, but only to the optimal subspace spanned by such $K$ lowest eigenstates.
In other words, the optimal  states $\lbrace \ket{\Psi_j(\bmtheta)} \rbrace$ at convergence
are not expected to be correct variational approximations of the targeted eigenstates of $\hat{H}$ but only mutually rotated versions of them, because $E_C^\bfw(\bmtheta)$ with equal weights is the Hamiltonian subtrace, which is invariant with respect to any rotation among the subspace states $\lbrace \ket{\Psi_j(\bmtheta)}\rbrace$.
Hence, an additional post-processing step is required (such as a classical diagonalization~\cite{parrish2019quantum} or additional rotations among the many-body states~\cite{yalouz2022analytical}) if one is interested in eigenproperties beyond the ensemble energy.
Nevertheless, in this work, we conclude that one
should always prefer the latter approaches, which have a very negligible computational overhead, rather than relying on ordered weights.

To assess the aforementioned statement, we will study two different problems: (1) the two-state many-body problem of the formaldimine molecule (SA-2-CAS(4,3)SCF level of theory) that exhibits strong vibronic coupling (around a conical intersection between the ground state and the first excited singlet)~\cite{formaldimine_85,
yalouz2021state,yalouz2022analytical},
and (2) the eight-state one-body problem corresponding to the eight restricted-Kohn-Sham doubly-occupied spatial-orbitals of a half-filled (16 electrons) linear chain of 16 hydrogen atoms, derived from quantum density functional theory (Q-DFT)~\cite{senjean2023towards}.

To study the convergence towards the targeted eigen-subspace, we compute the normalized trace of the Hamiltonian matrix in the subspace spanned by all the states of the ensemble $\lbrace \ket{\Psi_j(\bmtheta)} \rbrace_{j=0,\hdots,K-1}$, since the trace is
invariant with respect to any rotation among
these states, {\it i.e.},
\begin{eqnarray}
E_T^\bfw(\bmtheta) &=& \dfrac{1}{K}\sum_{j=0}^{K-1}
\bra{\Psi_j(\bmtheta)} \hat{H} \ket{\Psi_j(\bmtheta)}\nonumber \\
&=&  \dfrac{1}{K}\sum_{j=0}^{K-1} E_j(\bmtheta),
\end{eqnarray}
where the dependence on the weights $\bfw$
is implicit and comes from the fact that
the states $\lbrace \ket{\Psi_j(\bmtheta)} \rbrace$ are obtained through
the optimization of the weight-dependent ensemble energy $E_C^\bfw$ in Eq.~(\ref{eq:cost}).
In the
particular case of equi-weights,
$E_T^\bfw = E_C^\bfw$~\cite{fan1949theorem,gross1988rayleigh,ding2024ground}.
In this work, the equi-weights satisfy $w_j = 1/K$ and the optimal weights $w_j = (2K-1-2j)/K^2$ from Ref.~\citenum{ding2024ground} are used in the context of ensemble VQE, {\it i.e.},
using Eq.~(\ref{eq:cost}) as a cost function and preparing $\lbrace \ket{\Psi_j(\bmtheta)} = \hat{U}(\bmtheta)\ket{\Phi_j} \rbrace_{j=0,\hdots,K-1}$ on $K$ different circuits.
In the following, we replace the numerical notation by the alphabetical notation
to avoid any confusion, as in practice the ordering of the energies of the prepared states may not match the ordering of the weights.
Hence, $\lbrace j = 0, \hdots, K-1 \rbrace \rightarrow \lbrace j = A, \hdots, K \rbrace$ where any alphabetical state can possibly be associated with the ground state denoted by 0.

\begin{figure}
\centering
\includegraphics[width=\linewidth, keepaspectratio]{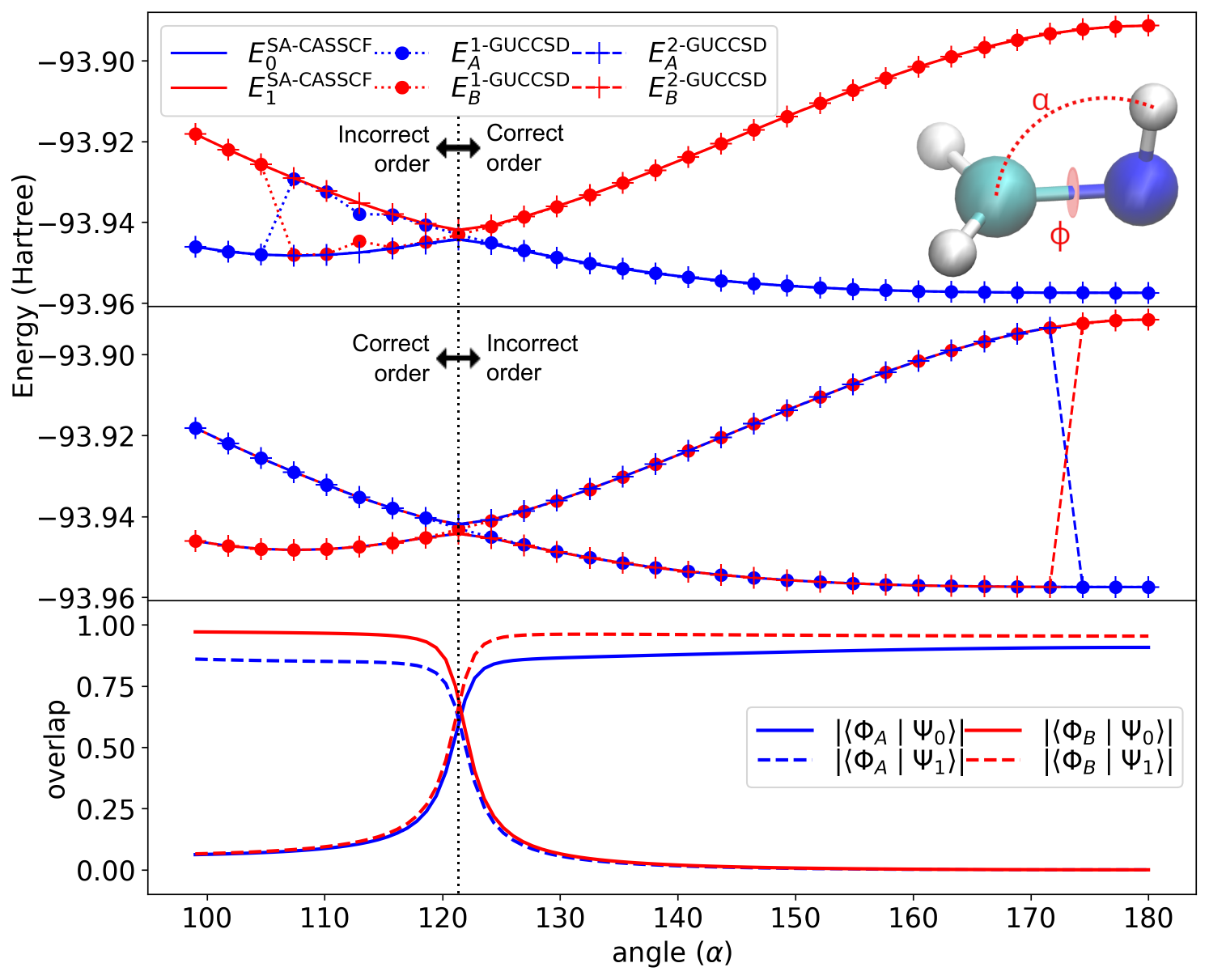}
\caption{
Ground and first-excited singlet-state potential energy surfaces (PESs) of the formaldimine molecule with respect to $\alpha$ at $\phi = 89^\circ$.
Top panel: 
$w_{A} > w_{B}$ with $w_{A} = 3/4$ and $w_{B} = 1/4$.
Middle panel: $w_{A} < w_{B}$ with $w_{A} = 1/4$ and $w_{B} = 3/4$.
Bottom panel: absolute values of the overlaps between the numerically exact eigenstates and the initial states of circuits $A$ and $B$.}
\label{fig:SAVQEPES}
\end{figure}
Let us begin our investigation with the two-state formaldimine case, where the Hamiltonian $\hat{H}$
is obtained from the state-average complete active space self-consistent field (SA-CASSCF) method with an active space of 4 electrons in 3 spatial orbitals,
with the same geometry as in Ref.~\citenum{yalouz2021state},
where we fix the dihedral angle $\phi = \widehat{\text{H--N--C--H}} = 89^\circ$ and vary the bending angle $\alpha = \widehat{\text{H--N--C}}$ from $99^\circ$ to $180^\circ$.
The initial states are the Hartree--Fock Slater determinant $\ket{\Phi_A}$
and the first-excited singlet configuration-state function ($n-\pi^*$ single excitation) $\ket{\Phi_B}$~\cite{yalouz2021state}.
In Fig.~\ref{fig:SAVQEPES}, we compare the energies of circuits $A$ and $B$
of the weighted-ensemble VQE,
using
$n = 1$ or $2$ repetitions of the generalized (disentangled) unitary coupled-cluster ansatz with single and double excitations~\cite{romero2018strategies} ($n$-GUCCSD) for $\hat{U}(\bmtheta)$ and
the gradient-based L-BFGS-G classical optimizer,
with the reference energies obtained from the ``exact'' diagonalization of $\hat{H}$,
derived from SA-CASSCF.

\begin{figure}
\centering
\includegraphics[width=\linewidth, keepaspectratio]{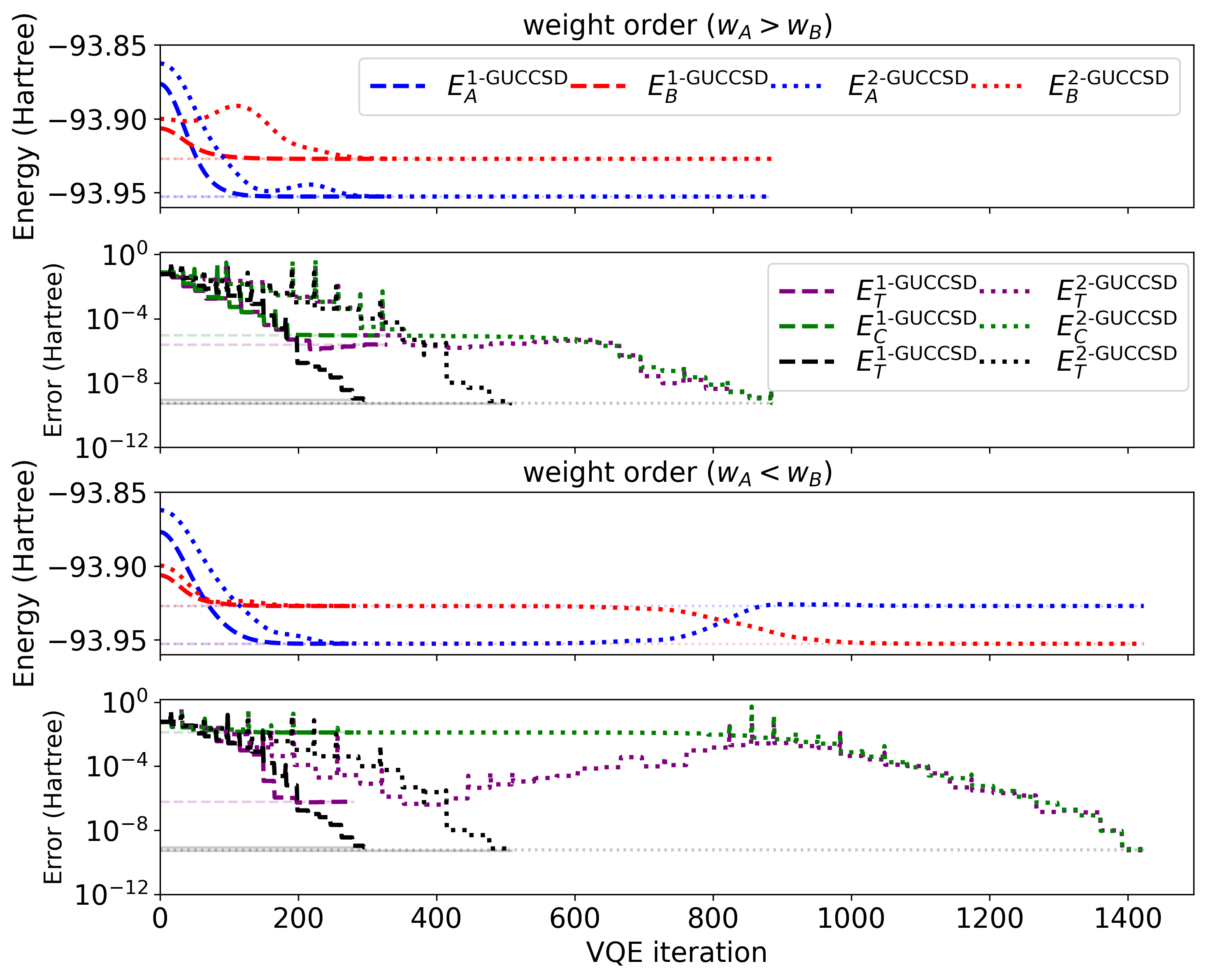}
\caption{Top panel:
behavior of $E_A$ and $E_B$ with respect to the number of VQE iterations until convergence (vanishing gradient of the cost function) for $\alpha = 138^\circ$ with $w_A > w_B$. Curves are smoothened using a Gaussian filter.
Horizontal lines (shaded colors) show the value of the last iteration of each curve (converged asymptote).
Middle-top panel: error in $E_C^\bfw$ and $E_T^\bfw$ using 1-GUCCSD (dashed lines) and 2-GUCCSD (dotted lines) in the weighted-ensemble (purple and green) and equi-ensemble (black) cases.
Middle-bottom and bottom panels: same as top and middle top panels but for $w_A < w_B$.}
\label{fig:SA_VQE_IT}
\end{figure}

Let us first focus on the right part of
the top panel ($w_A > w_B$)
and on the left part of the middle panel
($w_A < w_B$) of Fig.~\ref{fig:SAVQEPES}.
These regimes correspond to ideal scenarios where the order of the weights matches the values of the overlap between the initial states and the final states ({\it i.e.}, $w_A > w_B\leftrightarrow \psh{\Phi_A}{\Psi_0} > \psh{\Phi_B}{\Psi_0}$ and $\psh{\Phi_B}{\Psi_1} > \psh{\Phi_A}{\Psi_1}$) as well as the energy order of the final states after optimization with respect to the targeted eigenstates ($E_A \sim E_0 < E_B \sim E_1$).
Not surprisingly, both 1-GUCCSD and 2-GUCCSD manage to converge to properly ordered energies,
with an error of 10$^{-6}$ and $10^{-9}$ Hartree, and 300 and 900 SA-VQE iterations, respectively (see top panels of Fig.~\ref{fig:SA_VQE_IT} where, {\it e.g.},  $\alpha = 138^\circ$).
Turning to the incorrect order regime (left part of
the top panel and right part of the middle panel of Fig.~\ref{fig:SAVQEPES}),
the energy ordering is almost always violated 
when using 1-GUCCSD.
This leads to a huge error of $10^{-2}$ Hartree in the cost function (see bottom panel of Fig.~\ref{fig:SA_VQE_IT}).
However,
the error in the trace $E_T^\bfw$ remains of the order of $10^{-6}$ Hartree,
as in the ideal regime, thus
showing that 1-GUCCSD still manages to converge to the proper subspace by chance, but with a reverse energy order (local extremum of the cost function but same minimum of the trace).
Such a nice final outcome at some geometry cannot be guaranteed in a systematic manner everywhere. 
Indeed, we observe that for $\alpha = 113^\circ$, where the weights are also in the wrong order, we clearly do not obtain eigenstates.
This tells us that using ordered weights is an unpractical avenue, since this requires to know in advance the relative energy order of the final states started from a guess.
In contrast, an equi-weighted description does not suffer from such a predicament (see Fig.~\ref{fig:SA_VQE_IT}).

More surprisingly,
for $w_A < w_B$ (middle panel of Fig.~\ref{fig:SAVQEPES}),
the 1-GUCCSD PES
is regular and follows an ordering that seems to be dictated by the overlap between the initial and the final states (see bottom panel of Fig.~\ref{fig:SAVQEPES}).
This is reminiscent of the quasi-diabatic states that follow a maximal overlap criterion,
as shown for the equi-ensemble case using diabatic orbitals in a previous work~\cite{illesova2025transformation}.
This intriguing behavior,
which was not the expected one when using ordered weights, is certainly due to a strong local minimum of the cost function $E_C^\bfw$ being dominated by the global minimum of the trace contribution $E_T^\bfw$,
echoing also the convergence issues observed by Bierman and coworkers for weighted-ensemble VQE~\cite{bierman2022quantum}.
The reason why this local minimum seems to match the quasi-diabatic energies is an interesting open question that we plan to tackle in a future work.
This ordering issue is fixed everywhere by using the 2-GUCCSD ansatz
except for the last three points of the PESs of the middle panel of Fig.~\ref{fig:SAVQEPES},
certainly because the local minima sharpens
with the increase in overlap between the initial state and the final state, showing again how crucial the choice of the guess is.

To provide additional analysis and compare with the equi-ensemble case (for which individual-state energies are not meaningful), we provide
the error in $E_C^\bfw$ and $E^\bfw_T$ compared to exact diagonalization in
Fig.~\ref{fig:SA_VQE_IT}, using $w_A > w_B$ (first two top panels)
and $w_A < w_B$ (last two bottom panels)
for $\alpha = 138^\circ$,
which correspond to the correct and the incorrect orders, respectively.
Starting with the equi-ensemble case,
both 
1-GUCCSD and 2-GUCCSD converge to the exact eigen-subspace within $10^{-9}$ Hartree, requiring only around 300 and 500 iterations, respectively. 
Hence, the large computational overhead due to the use of 2-GUCCSD is completely unnecessary here.
In contrast, the use of ordered weights within 1-GUCCSD
leads to an error of $10^{-6}$ Hartree for the trace, and $10^{-5}$ Hartree for the cost function in the ideal case ($10^{-2}$ Hartree otherwise), with a fast convergence of 300 iterations as well.
To reduce the error for both the trace $E_T^\bfw$ and the cost function $E_C^\bfw$,
2-GUCCSD is required, thus increasing the computational complexity and the number of SA-VQE iterations to 900 with the correct ordering ($w_A > w_B$)
and to 1400 with the incorrect ordering ($w_A < w_B$).

Interestingly, the convergence of 2-GUCCSD first follows the same behaviour as 1-GUCCSD, {\it i.e.}, the error on the trace
goes down monotonically to $10^{-6}$ Hartree,
while the cost function follows the same path when correct ordering is used (middle-top panel) or gets stuck to a plateau with a large error if incorrect ordering is used (bottom panel).
But when the same accuracy than 1-GUCCSD is reached, 2-GUCCSD performs an additional step.
In the ideal case, it simply reduces the error on both the trace and the cost function to $10^{-9}$ Hartree (middle-top panel).
However, in the non-ideal case (bottom panel),
the error in the trace starts to increase again
in between 400 and 800 iterations while the cost function remains unchanged, showing that the classical optimizer tries to get out of the local minimum, hence deteriorating the eigen-subspace to achieve a better convergence on the cost function.
Indeed, after 800 iterations, the states start to switch so that both the errors on the cost function $E_C^\bfw$ and the trace
$E_T^\bfw$ reach convergence to $10^{-9}$ Hartree after 1400 iterations.

Note that every point of the PES exhibits a similar behavior (see Appendix~\ref{app:more_examples}).
Hence, enforcing an \emph{a priori} ordering
of the states makes the convergence much more difficult, since the classical optimizer
first ends up in a local minimum corresponding to the proper eigen-subspace (certainly according to a least-action principle as suggested in Ref.~\citenum{illesova2025transformation}) but needs to find the individual eigenenergies, as well as their proper ordering,
which first leads to a deterioration of the eigen-subspace and thus requires many more iterations.
These observations reinforce the conclusions of Ref.~\citenum{bierman2022quantum} regarding the slow convergence of weighted-ensemble VQE (see Fig.~9 of Ref.~\citenum{bierman2022quantum}, where
weighted-ensemble VQE optimizes each individual energy successively, starting from the ground-state one, instead of optimizing everything at once).

Interestingly enough, the difference between the equi-ensemble and
the 2-GUCCSD weigthed-ensemble results in the bottom panel of Fig.~\ref{fig:SA_VQE_IT} is reminiscent of the results obtained with the
variance-based VQE approach~\cite{zhang2022variational} (see Fig.~3 of Ref.~\citenum{zhang2022variational}), where the variance-based cost-function tends to converge directly to the nearest eigenstates rather than maintaining a prescribed ordering, at the cost of estimating the variance at each iteration.
In contrast, the equi-ensemble without variance will
not converge towards the nearest eigenstates,
but eventually towards the quasi-diabatic states as shown in Ref.~\citenum{illesova2025transformation}.
Note that the quasi-diabatic states are sometimes very similar to the eigenstates, as readily seen in Fig.~\ref{fig:SAVQEPES} except around the avoided crossing, as expected.


\begin{figure}
	\centering
	\includegraphics[width=\linewidth, keepaspectratio]{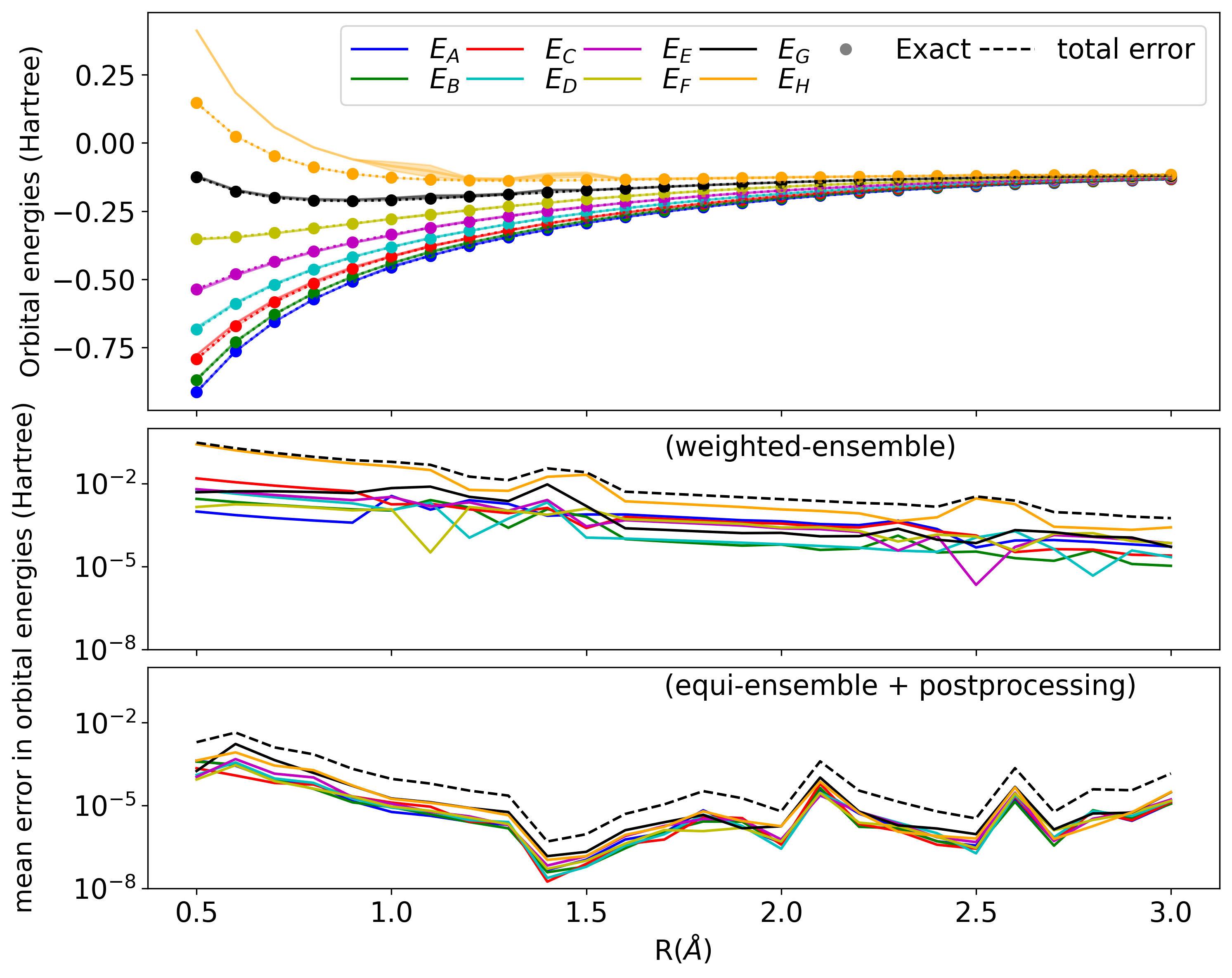}	
	\caption{Top panel:
	Mean values of the occupied orbital energies of H$_{16}$ over
    10 trials using random initial $\bmtheta$-parameters, within the optimally weighted-ensemble (full lines) and
    the equi-ensemble with classical postprocessing diagonalization (dashed lines). Shaded areas correspond to standard deviations (only visible for $E_H$ within the weighted-ensemble).
    `Exact' (dot markers) refers to the energies obtained by diagonalizing the Kohn--Sham Hamiltonian.
	Middle panel: mean error in each orbital energy obtained from the weighted-ensemble
    compared with exact diagonalization.
	Bottom panel: Same as middle panel but using the equi-ensemble followed by a postprocessing classical diagonalization.}
	\label{fig:Exc_H16}
\end{figure}
Let us now turn to another challenging example: an ensemble of eight one-body states
corresponding to the eight doubly-occupied restricted-Kohn--Sham spatial-orbitals of a linear chain of 16 equi-distant hydrogen atoms, derived from the first iteration of Q-DFT~\cite{senjean2023towards} ({\it i.e.}, the Kohn--Sham Hamiltonian $\hat{H}^{\rm KS}$, functional of the density, with the superposition of atomic density approximation for the density).
Q-DFT maps the non-interacting problem of
$N$ spatial-orbitals (within the restricted Kohn--Sham DFT formalism) onto a ``formally interacting'' $\log_2(N)$-qubit one.
The basis of orthogonalized atomic orbitals obtained from the L\"owdin symmetric orthonormalization~\cite{lowdin1950non} of the minimal STO-3G basis set is used, 
thus leading to $N = 16$ spatial orbitals mapped to 4 qubits (according to a binary numeral mapping).
For the SA-VQE part, we use the same formula for the optimal weights, the L-BFGS-B classical optimizer, and 10 layers of the linearly-entangled R$_y$CNOT hardware efficient ansatz~\cite{kandala2017hardware}
for $\hat{U}(\bmtheta)$.
See Appendix~\ref{app:savqe_qdft} for more details.

Fig.~\ref{fig:Exc_H16} exhibits the energies 
and errors
of the first eight eigenvalues of $\hat{H}^{\rm KS}$
with respect to the equi-distance $R$ between the hydrogen atoms,
obtained from the
weighted-ensemble and
equi-ensemble VQE.
As readily seen in the top panel of Fig.~\ref{fig:Exc_H16},
the low-lying excited-state energies
are relatively well reproduced compared to the highest one within the weighted-ensemble case,
thus showing a large and non-democratic error
(more than $10^{-1}$ Hartree for $R = 0.5$~\AA)
largely dominated by the highest-energy state associated to the lowest weight~\cite{bierman2022quantum},
as also shown in the middle panel.
In contrast, according to
the bottom panel of Fig.~\ref{fig:Exc_H16},
the
use of an equi-ensemble with classical postprocessing diagonalization of the Hamiltonian matrix -- which is not supposed to be diagonal in the basis of the equi-ensemble SA-VQE states --
reduces the error by two or more orders of magnitude for $R < 2$~\AA,
and around one order of magnitude for $R > 2$~\AA.
More importantly, the error does not seem
dominated by a given state and remains democratic,
which is a better feature.
Besides, the number of iterations required to reach convergence
is usually twice as lower as in the weighted-ensemble case (see Appendix~\ref{app:more_qdft}).
Hence, it is clear that the equi-ensemble
case is always performing much more accurately and efficiently than the weighted-ensemble one to achieve its target.
Regarding Q-DFT, one could argue that the number of states in the ensemble can be very large (as it corresponds to the number of occupied orbitals), and
that the post-processing classical diagonalization to get
the final eigenvalue will lead to a significant overhead, in contrast to the many-body electronic structure problem that usually focuses on a few low-lying excited states.
This is true if one wants to have access to the Kohn--Sham orbitals and Kohn--Sham orbital energies.
However, the total energy in DFT does not require the latter, but only the ensemble energy $E_T^\bfw$ of
the occupied subspace, as well as the electronic density obtained from it.
Hence, the equi-ensemble VQE can be used without any post-processing at all in the context of Q-DFT (see Appendix~\ref{app:savqe_qdft}).

To summarize, we have shown that the equi-ensemble VQE performs much better
than the weighted-ensemble VQE to
reach the proper low-lying excited-state eigensubspace.
It leads to more robust and unbiased optimization,
and always provides the correct eigen-subspace in contrast to the latter.
This has been shown
on the many-body electronic structure problem
exhibiting an avoided crossing, as
well as on the non-interacting Kohn--Sham
system in quantum density functional theory,
both representatives of typical systems for which
quantum computers are expected to be advantageous in the future.
Despite the fact that using an equi-ensemble
does not allow one to extract the eigenstates and
associated energies directly, so that
additional post-processing classical diagonalization~\cite{parrish2019quantum}
or post-variational rotations~\cite{yalouz2022analytical}
are required, our results
highlight that these post-processing
methods should always be preferred to weighted-ensemble VQE.
Besides, some cases do not even require the eigenstates, for instance when quasi-diabatic states are required~\cite{illesova2025transformation}
or when only a given eigen-subspace is targeted~\cite{senjean2023towards}.

\section*{Acknowledgement}

The authors thank Dr. Saad Yalouz
for fruitful discussions.
This research was funded in part by the
French National Research Agency (ANR) under the project ANR-23-CE29-0004-01
and under
the ``Investissements d'avenir'' program with the reference ANR-16-IDEX-0006.
This work benefited also from State support managed by the ANR under the France 2030 program, referenced by ANR-23-PETQ-0006.  Finally this work was also funded by Italian Government (Ministero dell'Università e della Ricerca, PRIN 2022 PNRR) -- cod.P2022SELA7: ''RECHARGE: monitoRing, tEsting, and CHaracterization of performAnce Regressions`` -- Decreto Direttoriale n. 1205 del 28/7/2023.

\bibliography{reference}

\appendix

\section{The many-body electronic structure problem of the Formaldimine molecule}
\label{app:formaldimine}

\subsection{The frozen-core Hamiltonian}
\label{app:frozen-core}

Within the Born--Oppenheimer approximation,
the time-independent and
non-relativistic electronic structure Hamiltonian
reads, under second quantization,
\begin{eqnarray}
    \hat{H}
= 
\sum_{pq} h_{pq} \hat{E}_{pq} + \dfrac{1}{2}
\sum_{pqrs}
g_{pqrs}\hat{e}_{pqrs},
\end{eqnarray}
where the one- and two-electron integrals are defined as (in real algebra)
\begin{eqnarray}
h_{pq} = \int \, \ddroit \bfr \phi_p(\bfr) \left( - \dfrac{1}{2}\nabla^2_\bfr + v_\textrm{ne}(\bfr) \right)
\phi_q(\bfr),
\end{eqnarray}
and
\begin{eqnarray}
g_{pqrs} = \iint \ddroit \bfr_1 \ddroit \bfr_2\, \dfrac{\phi_p(\bfr_1)\phi_r(\bfr_2)\phi_q(\bfr_1)\phi_s(\bfr_2)}{|\bfr_1 - \bfr_2|},
\end{eqnarray}
respectively,
where $\lbrace \phi_p \rbrace$ are the spin-restricted spatial molecular orbitals defining the (finite) basis set and $v_\textrm{ne}(\bfr)$ is the nucleus-electron potential.
The one- and two-body spin-free excitation operators are defined as $\hat{E}_{pq} = \sum_{\sigma} \hat{a}_{p\sigma}^\dagger \hat{a}_{q\sigma}$
and $ \hat{e}_{pqrs} = \sum_{\sigma,\tau} \hat{a}_{p\sigma}^\dagger \hat{a}_{r\tau}^\dagger \hat{a}_{s\tau}\hat{a}_{q\sigma}$,
where $\hat{a}_{p\sigma}^\dagger$ ($\hat{a}_{p\sigma}$) is the creation (annihilation) operator of an electron with spin $\sigma$ in spatial orbital $p$.

The formaldimine molecule has 16 electrons and a minimal basis of 13 core and valence atomic orbitals forming two core and six valence doubly-occupied spatial molecular orbitals, as well as five valence unoccupied spatial molecular orbitals as regards the spin-restricted closed-shell Hartree-Fock wavefunction.
For the larger cc-pVDZ basis considered in this work, the number of spatial molecular orbitals
is 43 (which would require
86 qubits
using the Jordan--Wigner mapping if one were to try a full CI description).
Due to the exponential increase of the size of the configuration space with respect to the number of molecular orbitals,
it is of common use to select only a restricted but relevant part of it 
by considering the active-space (or frozen-core) approximation where
the orbital space is separated into a set of frozen (doubly-occupied), active (varyingly-occupied), and virtual (unoccupied) orbitals.
Hence, every Slater determinant $|\Phi\rangle$ will always take the following antisymmetric form, 
\begin{equation}
    |\Phi\rangle = | \Phi_\text{core}\Phi_\text{active} \rangle,
    \label{eq:state}
\end{equation}
where $\Phi_{\rm core}$ represents the part of the determinant encoding the frozen-core orbitals of the system (always doubly-occupied) whereas $\Phi_{\rm active}$ is the part encoding the occupancy of the additional electrons in the active orbitals of the system.
The remaining virtual orbitals are always left unoccupied.
Using this definition,
we can write the
Hamiltonian into an effective form,
\begin{eqnarray}\label{eq:Ham_elec}
\bra{\Phi} \hat{H} \ket{\Phi} \equiv \bra{\Phi_{\rm active}} \hat{H}^{\rm FC} \ket{\Phi_{\rm active}},
\end{eqnarray}
with $\hat{H}^{\rm FC}$ the so-called ``frozen-core Hamiltonian'' defined as follows,
\begin{eqnarray}
\label{eq:Ham_elec}
\hat{H}^{\rm FC} = \hat{H}_\text{active} + E_\text{frozen}^\text{MF} + \mathcal{\hat{V}}.
\end{eqnarray}
Here, $\hat{H}_\text{active}$ is the Hamiltonian encoding the one- and two- body terms only acting in the active space,
\begin{equation}
\label{eq:AS_HAM}
\hat{H}_\text{active} = \sum_{tu}^\text{active} h_{tu} \hat{E}_{tu} + \sum_{tuvw}^\text{active} g_{tuvw} \hat{e}_{tuvw},
\end{equation}
where $t,u,v,w$ denote active space orbitals.
The second term $E_{\rm frozen}^{\rm MF}$ is a scalar representing the mean-field-like energy obtained from the frozen orbitals,
\begin{equation}
    E_\text{frozen}^\text{MF} = 2\sum_i^\text{frozen} h_{ii} + \sum_{ij}^\text{frozen} (2g_{iijj}- g_{ijji}),
    \label{eq:shift}
\end{equation}
and the third term
\begin{equation}
\label{eq:emb}
\mathcal{\hat{V}} = \sum_{tu}^\text{active} \mathcal{V}_{tu} \hat{E}_{tu} \text{, with } \mathcal{V}_{tu} =  \sum_i^\text{frozen} (2g_{tuii}- g_{tiiu} )
\end{equation}
represents an effective one-body potential, which encodes the interaction of the electrons of the active orbitals
with the ones of the frozen doubly-occupied orbitals.

Note that in a reduced configuration space,
the configuration-interaction method is no longer invariant under orbital rotations and the orbitals need to be re-optimized
(see Refs.~\cite{helgaker2014molecular,yalouz2021state}).
In this Letter, we employ
$\hat{H}^{\rm FC}$
using the converged orbitals of the SA-MCSCF method implemented in Psi4~\cite{PSI4},
using the
equi-ensemble for $S_0$ and $S_1$ and
the active-orbital subset $(\pi,n,\pi^*)$ within a CAS(4,3) description (complete active space with four electrons in three spin-restricted spatial orbitals),
which accounts for the relevant excitations and minimal static correlation of the system.

\subsection{The state-average variational quantum eigensolver}
\label{app:savqe}

Around the ground-state equilibrium geometry of formaldimine, the ground state is essentially a closed-shell Slater determinant of Hartree--Fock-type, while the first-excited singlet state is dominated by a singlet open-shell configuration-state function (CSF) of $n\pi^*$-type~\cite{illesova2025transformation}.
These
are exactly the initial states considered in this work.
While the Hartree--Fock state $\ket{\Phi_A}$ is easy to prepare within the Jordan--Wigner mapping,
the singlet open-shell CSF $\ket{\Phi_B}$ is slightly
more complicated; 
its circuit is provided in Ref.~\cite{yalouz2022analytical}.
Note, however, that the corresponding circuit depth is completely negligible in comparison to the ansatz.

Indeed, having prepared the two initial states,
we now apply the generalized unitary coupled-cluster ansatz with single and double fermionic excitations~\cite{romero2018strategies},
\begin{eqnarray}
    \hat{U}(\bmtheta) = e^{\hat{T}(\bmtheta) - \hat{T}^\dagger(\bmtheta)}
\end{eqnarray}
where
\begin{eqnarray}
    \hat{T}(\bmtheta) =
    \sum_{pq} \theta_{q}^p \hat{a}_{p}^\dagger \hat{a}_q + 
\sum_{pqrs}
\theta_{rs}^{pq} \hat{a}_{q}^\dagger \hat{a}_p^\dagger \hat{a}_s \hat{a}_r,
\label{eq:T_op}
\end{eqnarray}
and the indices run over every active orbitals,
hence the prefix `generalized'.
After the Jordan--Wigner encoding,
this ansatz reads
\begin{eqnarray}
    \hat{U}(\bmtheta) = e^{\sum_k \theta_k \hat{P}_k},
\end{eqnarray}
where $\hat{P}_k$ are Pauli words, the 
relation of which to Eq.~(\ref{eq:T_op}) can be found in Ref.~\cite{romero2018strategies}.
As such, an exponential operator is too complicated to
implement, and
it is of common use to
consider the `disentangled' (or Trotterized)
version of the ansatz,
\begin{eqnarray}
    \hat{U}(\bmtheta) \approx \prod_k e^{ \theta_k \hat{P}_k},
\end{eqnarray}
which is what is considered in this work.

Finally, once the states are prepared (ideally on a quantum computer), the expectation values
of the one- and two-body fermionic operators are
measured and gathered to compute the $\bfw$-weighted
cost function,
\begin{eqnarray} 
   E_C^\bfw(\bmtheta) &=&w_A \bra{\Phi_A}\hat{U}^\dagger(\bmtheta)
    \hat{H}^{\rm FC} \hat{U}(\bmtheta)\ket{\Phi_A} \nonumber \\
    &&+
     w_B \bra{\Phi_B}\hat{U}^\dagger(\bmtheta)
    \hat{H}^{\rm FC} \hat{U}(\bmtheta)\ket{\Phi_B},
    \label{eq:var_ES}
\end{eqnarray}
which upon minimization leads to the converged GUCCSD $\bmtheta$-parameters.
Note that the expectation value of the spin operator $\hat{S}^2$ was further added as a penalty term to Eq.~(\ref{eq:var_ES}) in order to favor singlet states ($\langle \hat{S}^2 \rangle = 0$), following the constrained VQE algorithm~\cite{ryabinkin2018constrained}.

\subsection{More examples on the formaldimine molecule}
\label{app:more_examples}

\begin{figure}
    \centering
    \includegraphics[width=1\linewidth]{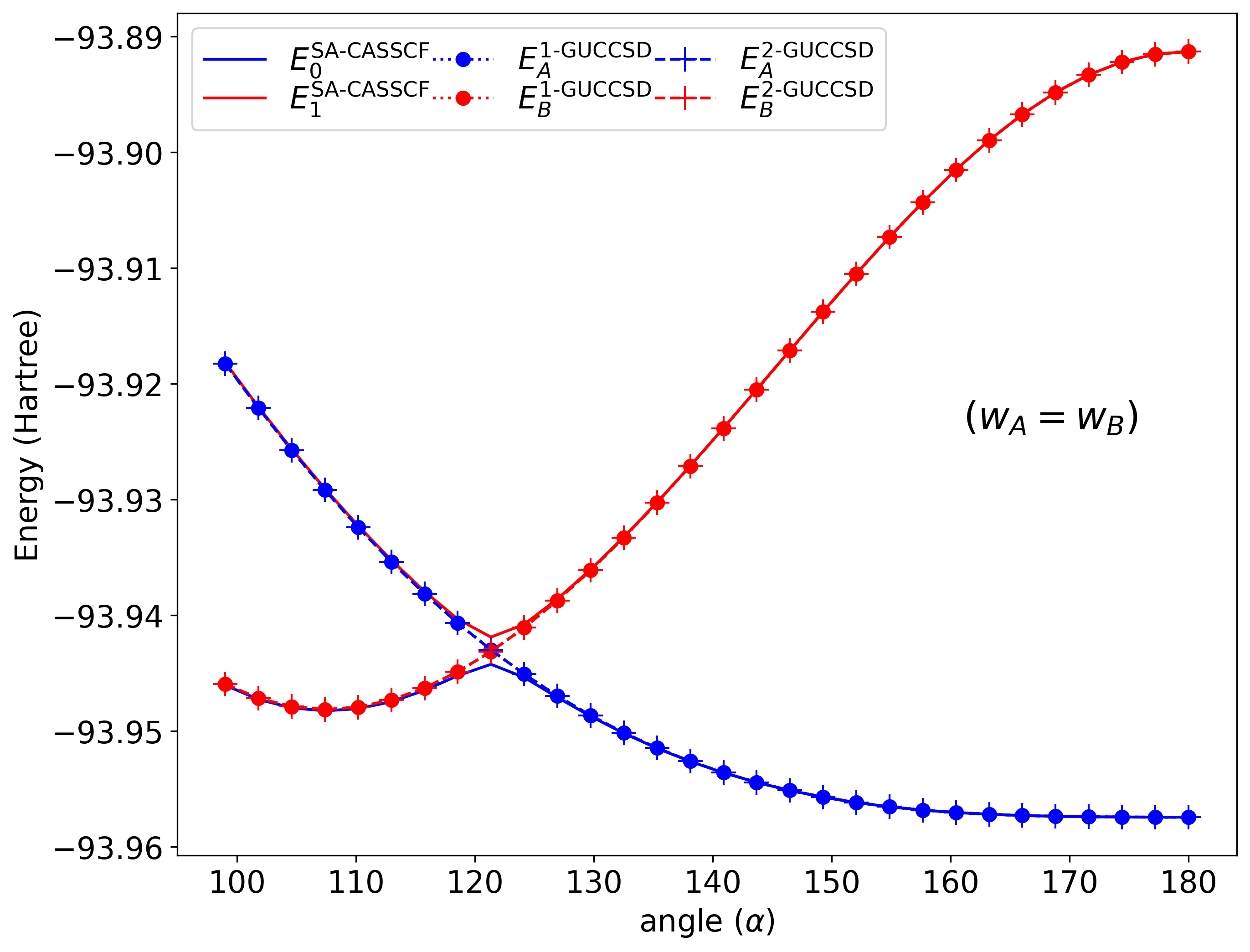}
    \caption{Potential energy surfaces of the formaldimine molecule with respect to $\alpha$ at $\phi = 89^\circ$ from the equi-ensemble VQE ($w_A = w_B$).} 
    \label{fig:pes_equi}
\end{figure}
Before presenting more results
to corroborate the discussion of the Letter,
we show in Fig.~\ref{fig:pes_equi}
the potential energy surfaces obtained using
the equi-ensemble VQE.
As readily seen in Fig.~\ref{fig:pes_equi},
the 1-GUCCSD energies are indistinguishable from the 2-GUCCSD ones, in contrast to
previously discussed results using the weighted-ensemble VQE.

Interestingly,
even though the equi-ensemble
energy is invariant
with respect to any rotation among the states $\lbrace \ket{\Psi_j(\bmtheta)}\rbrace$,
the fact that they both converge to the exact same values that seem
to correspond to quasi-diabatic energies
is not fortuitous
and certainly due to the least-action principle~\cite{illesova2025transformation}.
The 1-GUCCSD weighted-ensemble VQE
energies
in the middle panel of Fig.~1 of the Letter
are almost on top (although with a less accurate trace, {\it i.e.}, subspace) of the equi-ensemble ones,
but for wrong reasons as discussed in the Letter.
Note that diabatic orbitals were not used here to reach quasi-diabatic states, in contrast to Ref.~\cite{illesova2025transformation},
but they are essential for robustness
and to provide diabatic states for
any value of the dihedral angle $\phi$ (only $\phi=89^\circ$ is considered here).

\begin{figure}
\centering
\includegraphics[width=\linewidth, keepaspectratio]{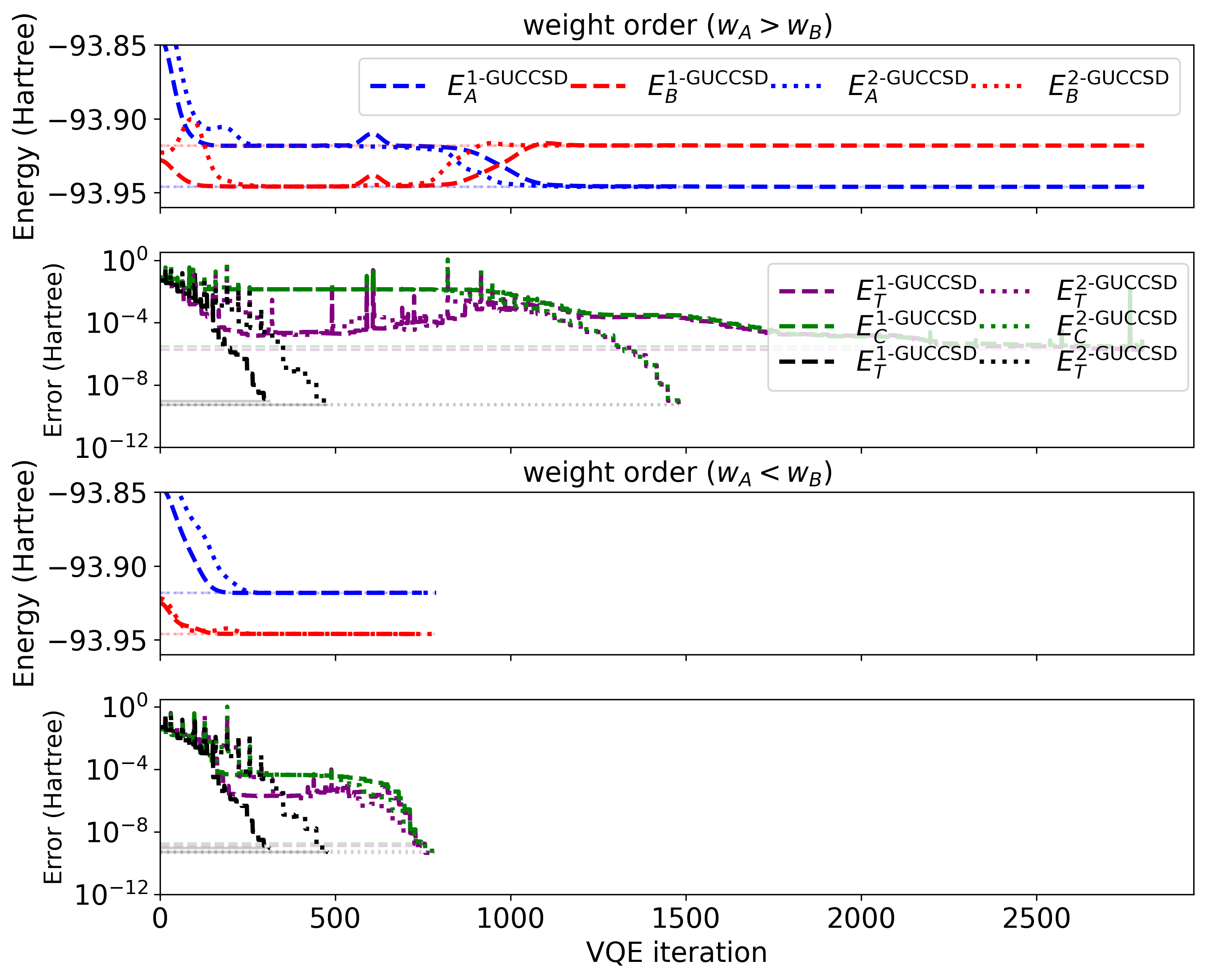}
\caption{Top panel:
$E_A$ and $E_B$ with respect to the number of VQE iterations for $\alpha = 99^\circ$ with $w_A > w_B$ (wrong initial order).
Middle-top panel: errors in $E_T^\bfw$ and $E_C^\bfw$ using 1-GUCCSD (dashed lines) and 2-GUCCSD (dotted lines) in the weighted-ensemble (purple and green) and equi-ensemble (black) case.
Middle-bottom and bottom panels: same as top and middle top panels but for $w_A < w_B$ (right initial order).}
\label{fig:convergence_plot_99}
\end{figure}
\begin{figure}
\centering
\includegraphics[width=\linewidth, keepaspectratio]{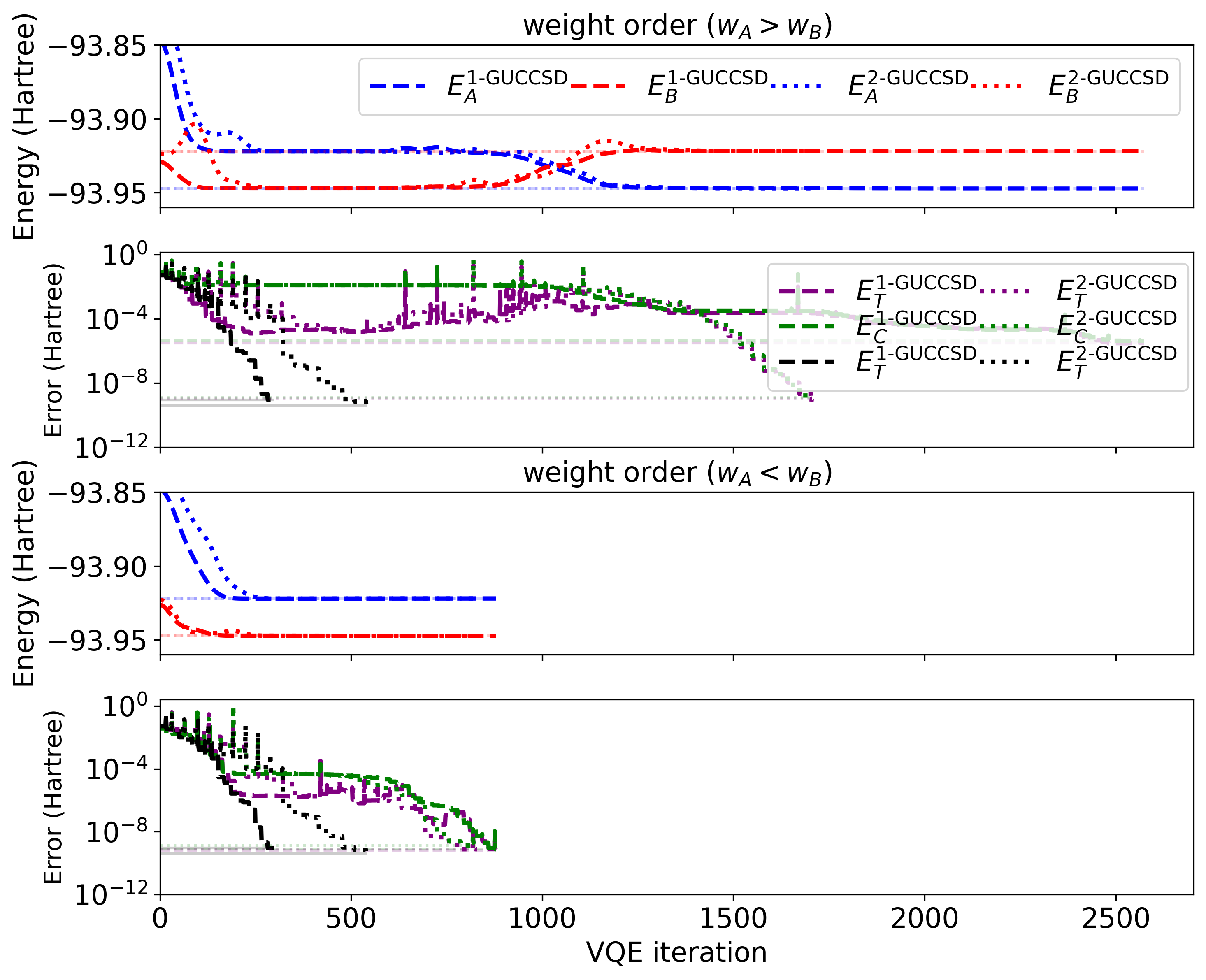}
\caption{Same as Fig.~\ref{fig:convergence_plot_99}
for $\alpha = 102^\circ$.}
\label{fig:convergence_plot_102}
\end{figure}
Let us now turn to the convergence of
the equi-ensemble and weighted-ensemble VQE
for several points of the PES.
In the Letter, $\alpha = 138^\circ$
was studied, and we show here that
the same analysis can be made for
any other points with other examples
at $\alpha = 99^\circ$,
$\alpha = 102^\circ$,
$\alpha = 113^\circ$,
$\alpha = 121^\circ$ (corresponding to the avoided crossing),
$\alpha = 160^\circ$, and
$\alpha = 180^\circ$.
As readily seen
in Figs.~\ref{fig:convergence_plot_99} and
\ref{fig:convergence_plot_102},
$w_A > w_B$ (top panels)
corresponds to the incorrect order, meaning
that $\psh{\Phi_A}{\Psi_0} < \psh{\Phi_A}{\Psi_1}$ and $\psh{\Phi_B}{\Psi_1} < \psh{\Phi_B}{\Psi_0}$.
Hence,
while the equi-ensemble VQE
converges really fast in 300 iterations,
the weighted-ensemble VQE takes
almost 3000 iterations
to converge only to $10^{-6}$ Hartree.
Note that in these cases,
1-GUCCSD finally manages to switch the states
in contrast to the case shown in the Letter ($\alpha = 138^\circ$),
which explains why the number of iterations is so high.
The swapping occurs at the same time
for 2-GUCCSD, but then converges much faster to $10^{-9}$ Hartree, in around 1500 iterations.
When the correct ordering is restored (bottom panels), every method manages
to reach $10^{-9}$ Hartree of accuracy
relatively fast (less than 1000 iterations)
although the convergence of the equi-ensemble
is at least twice faster.

\begin{figure}
\centering
\includegraphics[width=\linewidth, keepaspectratio]{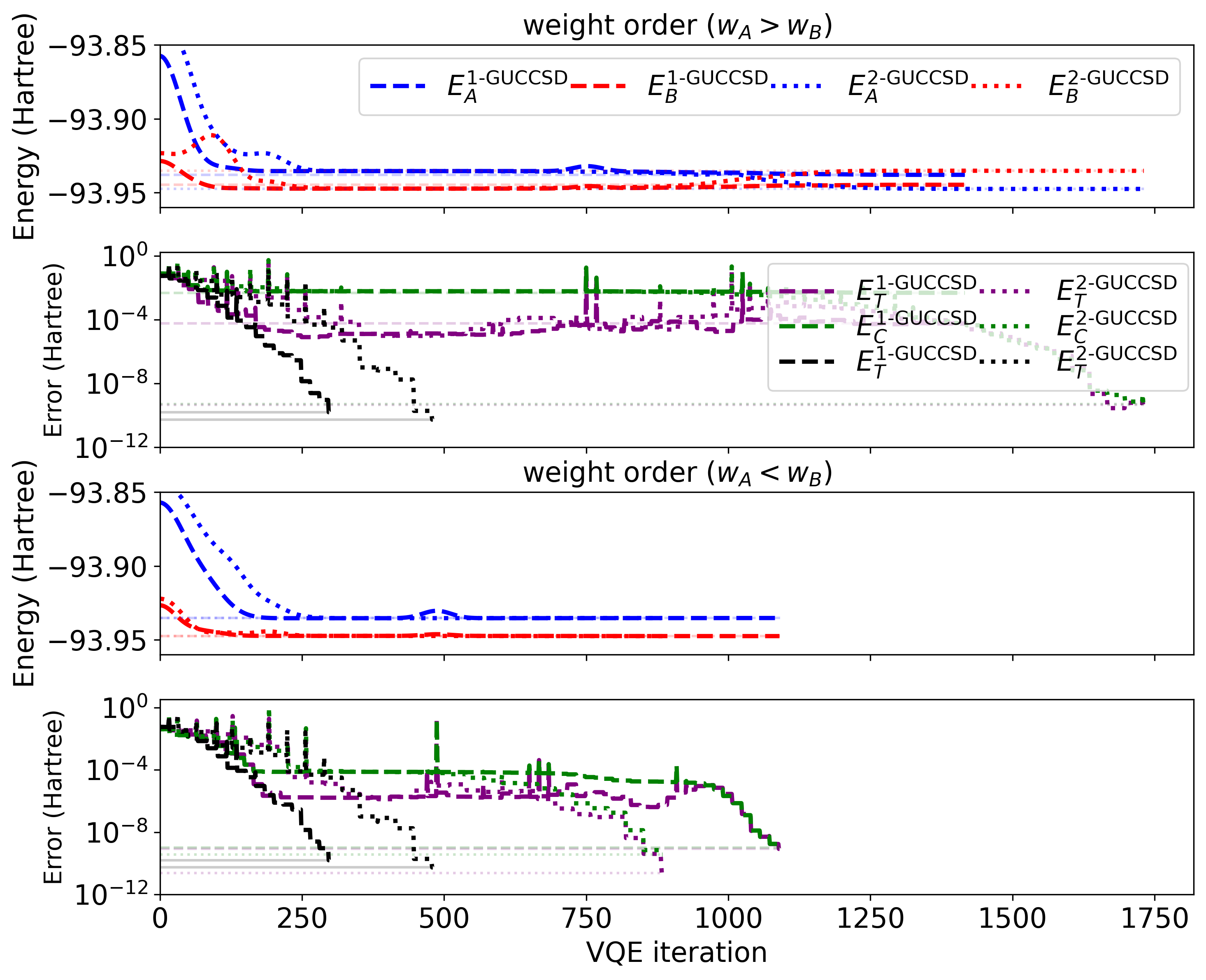}
\caption{Same as Fig.~\ref{fig:convergence_plot_99}
for $\alpha = 113^\circ$.}
\label{fig:convergence_plot_113}
\end{figure}
Turning to the most problematic point at
$\alpha = 113^\circ$ in Fig.~\ref{fig:convergence_plot_113} (top panels),
the error on the trace $E_T^\bfw$
is around $10^{-4}$ Hartree and
$10^{-2}$ Hartree for the cost function $E_C^\bfw$ using 1-GUCCSD, because the states have not been reordered accordingly during the optimization.
Looking at the middle-top panel of Fig.~\ref{fig:convergence_plot_113} (violet dashed line), we can see that
the trace was better converged during the optimization ({\it i.e.}, around $10^{-5}$ Hartree error at iteration 500) than at convergence.
Hence, minimizing slightly the cost function here (but not sufficiently to
reorder the states) has been done
at the expense of increasing the trace and thus deteriorating the target subspace.
This behavior is a striking signature of what 
can badly happen with the weighted-ensemble VQE when
ordered weights are not properly assigned to the initial states (which cannot be predicted in advance in general).
Again, within the correct ordering (bottom panels), everything converges as expected, although 2-GUCCSD is,  once more, converging faster than 1-GUCCSD for the weighted-ensemble VQE.
\begin{figure}
\centering
\includegraphics[width=\linewidth, keepaspectratio]{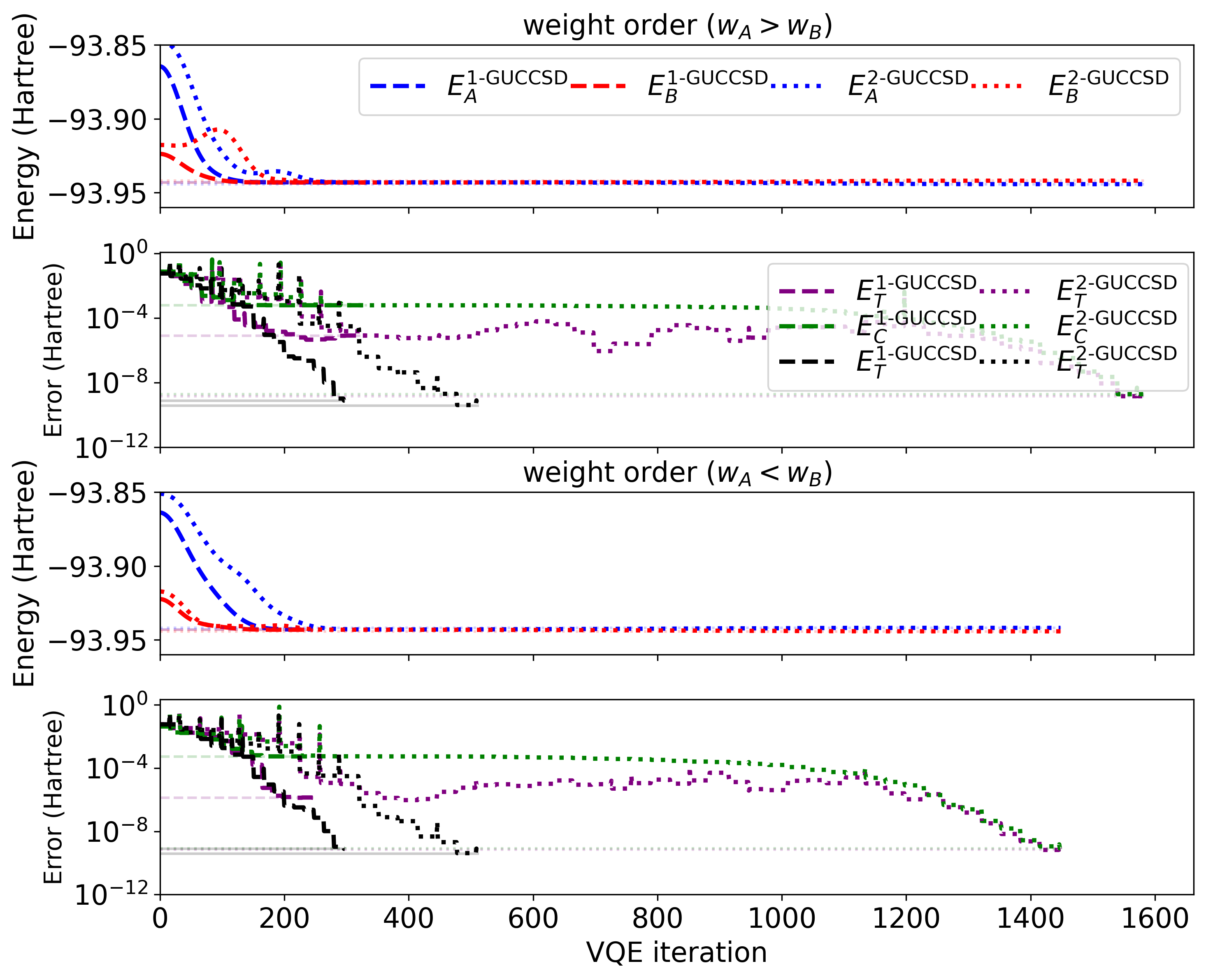}
\caption{Same as Fig.~\ref{fig:convergence_plot_99}
for $\alpha = 121^\circ$.}
\label{fig:convergence_plot_121}
\end{figure}

Another particular point one can look at is
the one corresponding to the avoided crossing, see Fig.~\ref{fig:convergence_plot_121}.
In such a case, no
particular changes are detected between
the top panels and the bottom panels.
Indeed, while the equi-ensemble VQE
still converges as easily as in any other case, the weighted-ensemble VQE
does not manage to reorder the states when using 1-GUCCSD, and seems to have difficulty ({\it i.e.}, requires many iterations) to 
do so using 2-GUCCSD.
This is because the avoided crossing corresponds to a geometry exhibiting
strong intrastate mixing, and where
the overlap between the eigenstates and
the initial states are almost equal,
{\it i.e.}, $\psh{\Phi_A}{\Psi_0} \sim \psh{\Phi_A}{\Psi_1}$
and $\psh{\Phi_B}{\Psi_0} \sim \psh{\Phi_B}{\Psi_1}$ (see Fig.~2 of the Letter).

\begin{figure}
\centering
\includegraphics[width=\linewidth, keepaspectratio]{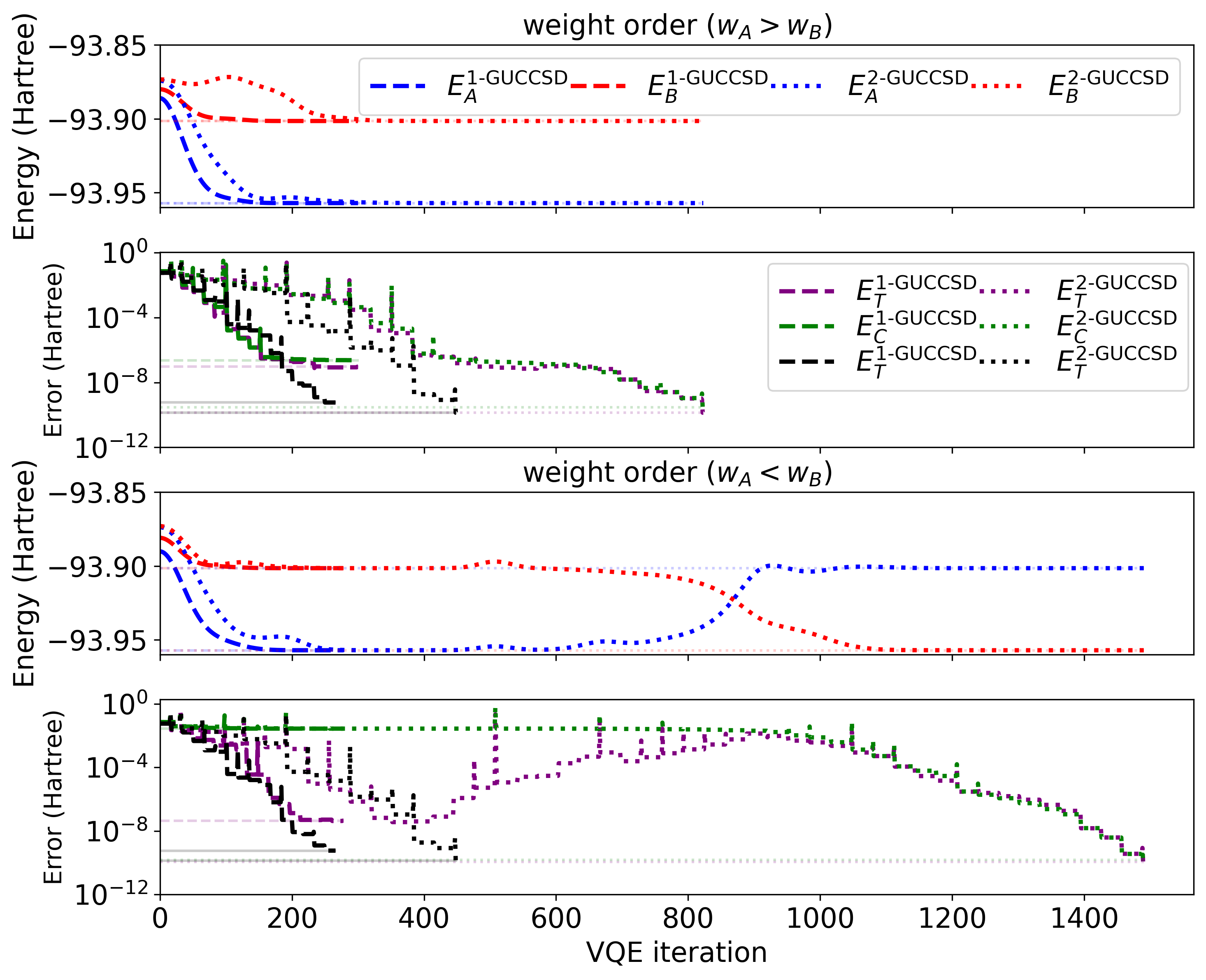}
\caption{Same as Fig.~\ref{fig:convergence_plot_99}
for $\alpha = 160^\circ$.}
\label{fig:convergence_plot_160}
\end{figure}
\begin{figure}
\centering
\includegraphics[width=\linewidth, keepaspectratio]{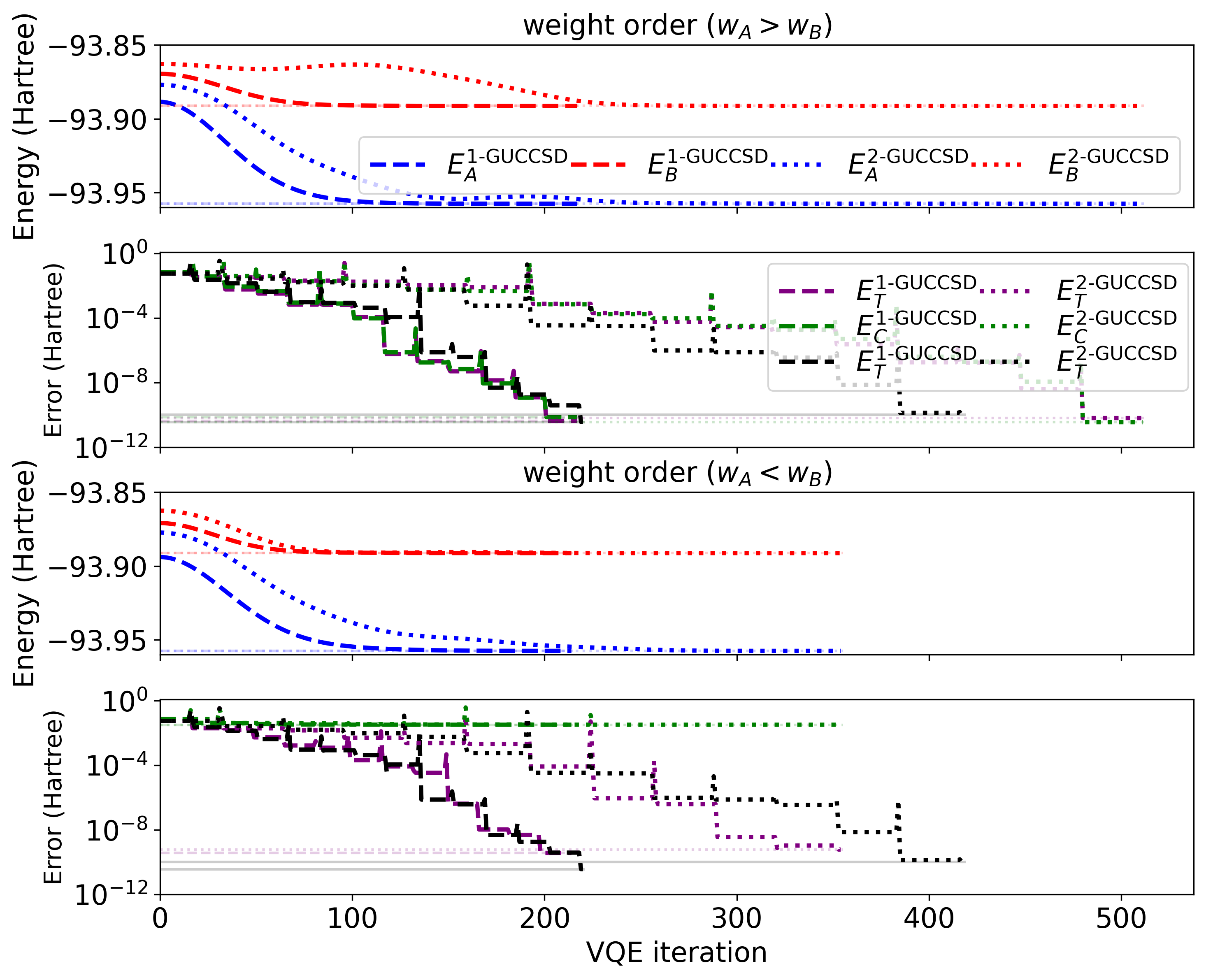}
\caption{Same as Fig.~\ref{fig:convergence_plot_99}
for $\alpha = 180^\circ$.}
\label{fig:convergence_plot_180}
\end{figure}
For $\alpha = 160^\circ$
in Fig.~\ref{fig:convergence_plot_160},
the exact same discussion as for $\alpha = 138^\circ$ holds.
Finally, for $\alpha = 180^\circ$ in Fig.~\ref{fig:convergence_plot_180},
no reordering of the states occurs regardless of the ordering of the weights.
In any case, the trace $E_T^\bfw$ converges with an error of $10^{-10}$ Hartree, showing that the proper eigenstates are actually obtained,
even though the correct ordering is sometimes not fulfilled (see bottom panel with the error of $10^{-2}$ Hartree on $E_C^\bfw$).
This can again be rationalized with 
the very high overlap values
$\psh{\Phi_A}{\Psi_0}$ and $\psh{\Phi_B}{\Psi_1}$,
while the crossed ones are zero there due to symmetry,
and because having the ensemble energy of the
wrongly-ordered eigenenergies also
corresponds to an extremum of the cost function.


\section{The one-body non-interacting Kohn--Sham problem for the 16-hydrogen chain}
\label{app:dft}

\subsection{The non-interacting Kohn--Sham Hamiltonian}
\label{app:ksdft_hamiltonian}

Kohn--Sham (KS) density functional theory relies on the transformation of the complex many-body electronic
structure problem into a non-interacting electronic density-dependent one-body effective problem~\cite{hohenberg1964electron,kohn1965ks}.
More precisely, the ground-state energy and ground-state electronic density of a given complex system can be obtained exactly by solving the KS equation~\cite{kohn1965ks}:
\begin{eqnarray}\label{eq:KS_eq_1}
\left( \hat{T} +  \hat{v}^{\rm KS}[n] \right) \ket{\Phi^{\rm KS}} = \mathcal{E}^{\rm KS}
\ket{\Phi^{\rm KS}},
\end{eqnarray}
where $\hat{T}$ is the kinetic energy operator, $\ket{\Phi^{\rm KS}}$ is a single Slater determinant
and $\hat{v}^{\rm KS}[n] = \hat{v}^{\rm ext} + \hat{v}^{\rm H}[n] + \hat{v}^{\rm xc}[n]$ is the density-dependent KS potential operator. 
It contains the external potential $\hat{v}^{\rm ext}(\bfr)$, \textit{i.e.}, the ion-electron interaction, the trivial Hartree potential $\hat{v}^{\rm H}[n](\bfr)$, and finally the so-called exchange and correlation (XC) potential $\hat{v}^{\rm xc}[n](\bfr)$, which implicitly contains all non-trivial contributions arising from the electron-electron interaction.
The KS equation~(\ref{eq:KS_eq_1}) is solved self-consistently 
by plugging the newly calculated electronic density
back into the KS potential, iteratively until convergence is reached.
At convergence, provided that the exact XC functional is known, the
non-interacting auxiliary KS system density is equal to the ground-state density of the physical system. 
The total energy, functional of the density,
is also recovered according to the following expression,
\begin{eqnarray}\label{eq:E0_KS}
E_0 &=& \mathcal{E}^{\rm KS} + E_{\rm Hxc}[n^{\Phi^{\rm KS}}] - 
\int \bfv^{\rm Hxc}[n^{\Phi^{\rm KS}}](\bfr) \,n^{\Phi^{\rm KS}}(\bfr) \, {\rm d} \bfr,\nonumber\\
\end{eqnarray}
where
\begin{eqnarray}
n^{\Phi^{\rm KS}}(\bfr) &=&
2 \sum_{k=1}^{N_{\rm occ}} | \varphi_k^{\rm KS}(\bfr) |^2 \nonumber \\
&=&
2\sum_{k=1}^{N_{\rm occ}} \psh{\bfr}{\varphi_k^{\rm KS}}\psh{\varphi_k^{\rm KS}}{\bfr},\label{eq:density}
\end{eqnarray}
and
\begin{eqnarray}
    \mathcal{E}^{\rm KS} = 2\sum_{k=1}^{N_{\rm occ}} \varepsilon_k^{\rm KS},
    \label{eq:energy_KS}
\end{eqnarray}
with
$\ket{\varphi_k^{\rm KS}}$
denoting the $k$-th KS spatial-orbital in $\ket{\Phi^{\rm KS}}$
and $\varepsilon_k^{\rm KS}$ its associated energy.
$N_{\rm occ}$ corresponds to the number of occupied spatial orbitals that is equal to half the number of electrons.

In this work, we do not aim at achieving self-consistency, and we simply extract the
KS Hamiltonian at first iteration,
{\it i.e.}, using the
sum of atomic densities for $n(\bfr)$
and the local spin density functional SVWN
for $\hat{v}^{\rm xc}[n]$.

\subsection{The state-average variational quantum eigensolver}
\label{app:savqe_qdft}

Considering the non-interacting KS Hamiltonian (Eq.~\ref{eq:KS_eq_1})
projected onto a finite computational basis set $\left\{\chi_{i}(\bfr)\right\}$ composed of $N$ spatial-orbitals,
\begin{eqnarray}\label{eq:H_mono}
    \hat{h}^{\rm KS} &= & \sum_{i=1,j=1}^{N}h_{ij}^{\rm KS}  \sum_\sigma   \left( \hat{c}^\dagger_{i\sigma} \hat{c}_{j\sigma} + \textrm{h.c.} \right),
  \end{eqnarray}
where $\hat{c}^\dagger_{i\sigma}$ ($\hat{c}_{i\sigma}$) refers to the creation (annihilation) operator of an electron of spin $\sigma={\uparrow,\downarrow}$ in the spatial-orbital $\chi_{i}(\bfr)$, respectively.  
The $h_{ij}^{\rm KS}$ matrix elements (integrals) involve both the kinetic contributions and the KS potential.  
We then map
the matrix representation of
$\hat{h}^{\rm KS}$, of dimension $N\times N$, onto
$M = \log_2(N)$ qubits following 
the binary mapping as suggested in Ref.~\citenum{senjean2023towards},
{\it i.e.},
each spatial-orbital
is encoded as a given bitstring of the $\log_2(N)$-qubit system.
The operators are
also transformed into
Pauli strings
following Refs.~\citenum{senjean2023towards} or \citenum{knapik2025quantum},
such that one can solve
Eq.~\ref{eq:KS_eq_1}
on a quantum computer
to extract the KS orbital and orbital energies
$\lbrace \ket{\varphi^{\rm KS}_k} \rbrace$ and
$\lbrace \varepsilon_k^{\rm KS} \rbrace$, respectively.
We refer to this method as Quantum Density Functional Theory (QDFT)~\cite{senjean2023towards}.

We employ the ensemble VQE algorithm
to extract the first $k=1,\hdots, N_{\rm occ}$ excited states and excited-state energies $\ket{\varphi^{\rm KS}_k}$ and
$\varepsilon_k^{\rm KS}$, respectively.
Hence, an ensemble of $N_{\rm occ} = 8$ states
are required for the hydrogen chain composed of 16 hydrogen atoms (16 electrons).
The hardware efficient ansatz, the $R_y$CNOT ansatz, is used for $\hat{U}(\bmtheta)$
and reads~\cite{kandala2017hardware}
\begin{eqnarray}\label{eq:Ry_ansatz}
\hat{U}(\bmtheta) &=& \prod_{m=1}^M R_{y,m}(\theta_{m_y}^0)\prod_{n=1}^{{N_L}} \hat{U}_n^{\rm ENT}(\bmtheta^n)
\end{eqnarray}
for a number of layers ${N_L}$ and a number of qubits $M$,
with the entanglement unitary blocks
\begin{eqnarray}\label{eq:entanglement}
\hat{U}_n^{\rm ENT}(\bmtheta^n) = 
\prod_{m=1}^{M-1} {\rm CNOT}_{m(m+1)} \prod_{m=1}^M R_{y,m}(\theta_{m}^n).
\end{eqnarray}
10 layers are considered in this Letter.
Just as for the formaldimine molecule,
one has to choose initial states
on which the transformation
$\hat{U}(\bmtheta)$
is applied.
According to Ref.~\citenum{senjean2023towards},
we choose the first states of the computational basis, \textit{i.e.}, $\lbrace \ket{ \phi_k} \rbrace = \lbrace 
\ket{0000},
\ket{0001}, 
\ket{0010},
\ket{0011},
\ket{0100},
\ket{0101},
\ket{0110},
\ket{0111}
\rbrace$,
that are very easy to prepare using maximum three $X$ gates.

Interestingly, the key quantities to be computed within QDFT are Eqs.~\ref{eq:density} and (\ref{eq:energy_KS}), which are both invariant with respect
to any rotation between the KS orbitals within the occupied subspace.
Indeed, Eq.~(\ref{eq:energy_KS}) is the
trace
of the Hamiltonian matrix, 
equivalent to $E_T^\bfw$ defined in the Letter.
For the density
in Eq.~(\ref{eq:density}),
applying a unitary matrix transformation ${\bf U}$
to the occupied orbitals of the KS Slater determinant
does not change the determinant (up to the sign
of the determinant of ${\bf U}$).
The associated idempotent one-body reduced density matrix
${\bm \gamma}^{\rm KS}$ is the identity
in the occupied KS orbital subspace (with a pre-factor 2 as we work with doubly-occupied spatial molecular orbitals, ${\bm \gamma}^{\rm KS}_{\rm occ} = 2{\bm I}$) and
0 elsewhere,
so that applying the unitary ${\bf U}$
to rotate the occupied orbitals
leads to the new density matrix
\begin{eqnarray}
    {\bm \lambda}_{\rm occ} &=& {\bf U}^\dagger {\bm \gamma}^{\rm KS}_{\rm occ} {\bf U} \nonumber \\
  &=&  2 {\bf U}^\dagger {\bm I} {\bf U} \nonumber \\
  &=& 2 {\bm I} \nonumber \\
  &=& {\bm \gamma}^{\rm KS}_{\rm occ},
\end{eqnarray}
where ${\bf U}^\dagger {\bf U} = {\bm I}$.
Hence,
the
equi-ensemble VQE is sufficient to solve
the non-interacting KS problem,
and there is no interest in considering
a weighted ensemble.

\subsection{More results on Quantum Density Functional Theory}
\label{app:more_qdft}

A detailed statistical analysis was performed, aiming to decide whether the improvements of equi-ensemble VQE are significant, i.e., whether the improvement can be expected in general or is only present occasionally. For this conclusion, we checked the normality of differences between $E_T^\bfw$ of equi-ensemble and weighted ensemble via the Shapiro-Wilk test~\cite{shapiro} with $p$-value equal to 0.05, same as for all the following tests. 
As can be seen from Tab.~\ref{tab:shapiro}, we are not sure about the normality of the 3 differences (in bold). 
Thus, we chose the non-parametric Wilcoxon Signed-rank test~\cite{conover1999practical} for comparison of the errors in the rest of this paper.  

For the comparison itself, we proceeded to compare both approaches via a) area under the curve~(AUC) as summary statistics and b) point-wise. In the former case, with AUCs listed in Tab.~\ref{tab:auc}, the Wilcoxon test's statistics is equal to 0 and its $p$-value to 0.001953125, thus deciding that the difference is significant and moreover, that all the differences are in the same direction, i.e., that the weighted ensemble demonstrates higher errors every time, when summarized via AUC. 

For the latter, point-wise analysis, we can see the results in Tab.~\ref{tab:pointwise}. Due to a large number of comparisons, we also adopted Benjamini-Hochberg correction~\cite{benjamini} to account for possible \textit{false discovery rate}. After this procedure, it can be observed that in 24 out of 26 values of $R$ the difference is significant, favoring the equi-ensemble approach again. Moreover, based on the listed statistics value, in 23 cases all the differences were in the same direction.

\begin{figure}
\centering
\includegraphics[width=\linewidth, keepaspectratio]{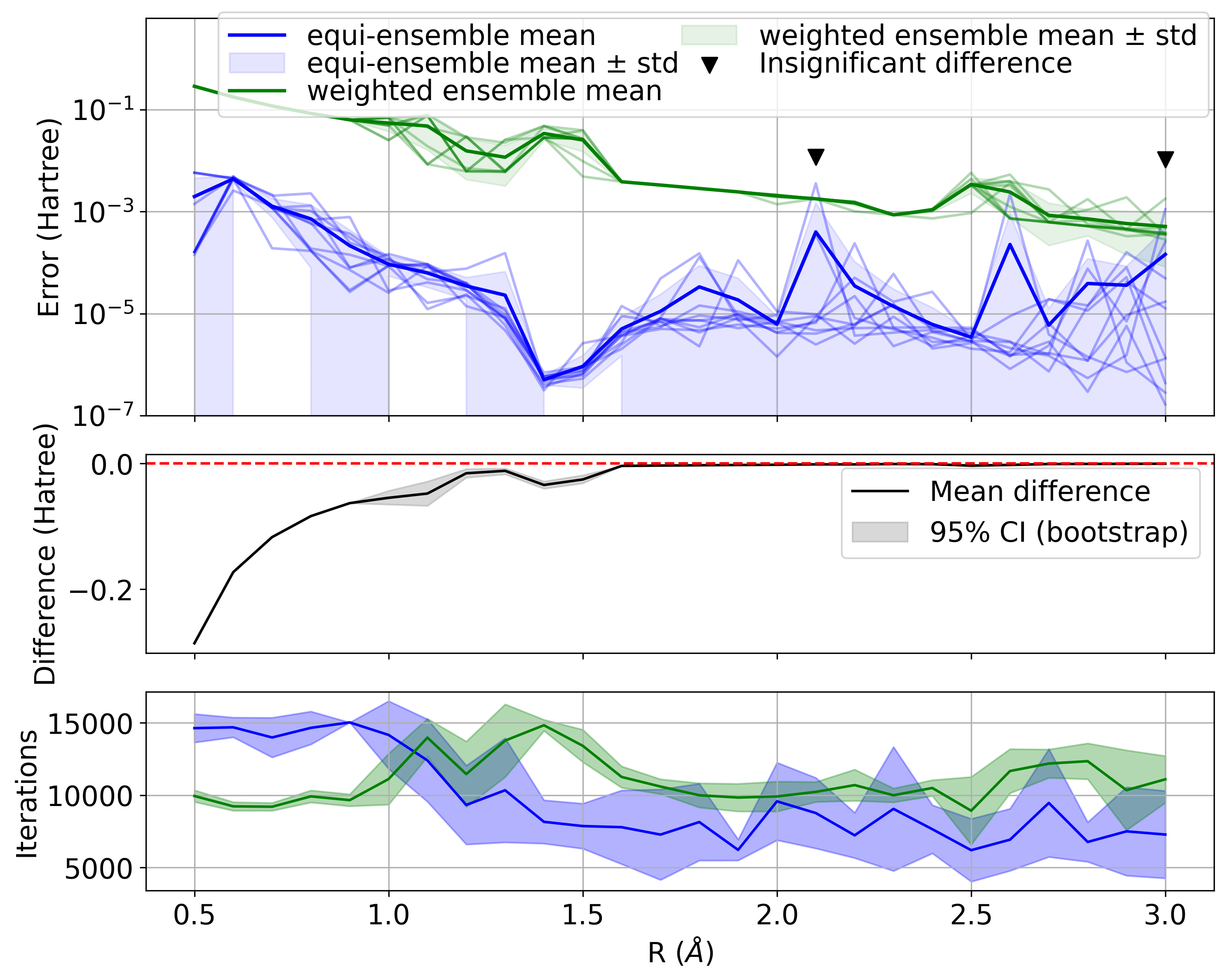}
\caption{10 trials per value of $R$ were considered, using randomly sampled initial values of the variational $\bm \theta$-parameters. Lines correspond to the mean values over the 10 trials, and shaded areas represent the corresponding standard deviations.
Top panel: error in the trace $E_T^\bfw(\bmtheta)$. Middle panel: Visualization of mean difference (between $E_T^\bfw(\bmtheta)$ of equi-ensemble and weighted ensemble) and the bootstrap confidence band.
Bottom panel: number of ensemble VQE iterations to convergence.
}
\label{fig:convergence_QDFT}
\end{figure}
For illustration purposes, the plot of both equi-ensemble and weighted ensemble errors is provided in Fig.~\ref{fig:convergence_QDFT}, together with their standard deviations and marks of the points, where the difference between both approaches was point-wise insignificant. It is joined together with the plot of VQE iterations necessary until convergence and with the plot of both mean difference between both approaches and the bootstrap confidence band~\cite{bootstrap} computed at 95\% confidence level and with 2000 samples, showing us the area, where 95\% of curves would be positioned, having made more runs.

In terms of computational efficiency, for small interatomic distances $0.5-1.2$~\AA,
the equi-ensemble VQE requires more iterations than the weighted-ensemble, which is
unusual and differs from what we have seen for the formaldimine molecule previously.
However, this range corresponds to where the use of ordered weights gives the largest error of around $10^{-1}$ Hartree on $E_T^\bfw$,
thus possibly ending in a local minimum that
does not affect the equi-ensemble case, which achieves much better accuracy albeit with more iterations.
Beyond this range, the equi-ensemble VQE
converges around twice as fast as the weighted-ensemble VQE together with better accuracy.

\begin{table}[!h]
\centering
\begin{tabular}{r c r c r c r c}
\hline
$R$ & $p$-value & $R$ & $p$-value & $R$ & $p$-value & $R$ & $p$-value \\
\hline
0.5  & 2.99e-04 &  0.6  & 2.26e-05 &  0.7  & 2.34e-02 &  0.8  & 2.53e-02 \\
0.9  & 3.61e-02 &  1.0  & 8.26e-03 &  1.1  & 6.20e-03 &  1.2  & 2.82e-04 \\
1.3  & 1.22e-04 & 1.4 & 6.82e-05 & 1.5 & 2.40e-02 & 1.6 & \textbf{4.60e-01} \\
1.7 & \textbf{1.12e-01} & 1.8 & 8.06e-03 & 1.9 & 3.44e-04 & 2.0 & 5.09e-07 \\
2.1 & 4.11e-07 & 2.2 & 1.23e-04 & 2.3 & 1.27e-04 & 2.4 & 4.54e-07 \\
2.5 & 6.72e-03 & 2.6 & \textbf{2.19e-01} & 2.7 & 2.68e-07 & 2.8 & 2.47e-03 \\
2.9 & 3.73e-06 & 3.0 & 3.27e-03 &    &         &    &         \\
\hline
\end{tabular}
\caption{Results of the Shapiro-Wilk test of differences between equi-ensemble and weighted-ensemble applied per value of $R$ (\AA).}
\label{tab:shapiro}
\end{table}

\begin{table}[!h]
\centering
\begin{tabular}{r c c}
\hline
Trial & Equi-ensemble & Weighted-ensemble \\
\hline
1  & 0.0154 & 0.890 \\
2  & 0.00593 & 0.962 \\
3  & 0.00916 & 0.953 \\
4  & 0.00661 & 1.016 \\
5  & 0.0133 & 0.922 \\
6  & 0.0130 & 0.945 \\
7  & 0.00669 & 0.929 \\
8  & 0.0111 & 0.918 \\
9  & 0.0111 & 1.029 \\
10 & 0.00565 & 0.924 \\
\hline
\end{tabular}
\caption{AUC values of each trial for both equi-ensemble and weighted-ensemble VQE.}
\label{tab:auc}
\end{table}

\begin{table}[!h]
\centering
\begin{tabular}{r c c c | r c c c}
\hline
$R$ & Stat. & $p$-value & BH & $R$ & Stat. & $p$-value & BH \\
\hline
0.5  & 0  & 0.00195 & 0.00221 & 1.8 & 0  & 0.00195 & 0.00221 \\
0.6  & 0  & 0.00195 & 0.00221 & 1.9 & 0  & 0.00195 & 0.00221 \\
0.7  & 0  & 0.00195 & 0.00221 & 2.0 & 0  & 0.00195 & 0.00221 \\
0.8  & 0  & 0.00195 & 0.00221 & 2.1 & \textbf{10} & \textbf{0.08398} & \textbf{0.08398} \\
0.9  & 0  & 0.00195 & 0.00221 & 2.2 & 0  & 0.00195 & 0.00221 \\
1.0  & 0  & 0.00195 & 0.00221 & 2.3 & 0  & 0.00195 & 0.00221 \\
1.1  & 0  & 0.00195 & 0.00221 & 2.4 & 0  & 0.00195 & 0.00221 \\
1.2  & 0  & 0.00195 & 0.00221 & 2.5 & 0  & 0.00195 & 0.00221 \\
1.3  & 0  & 0.00195 & 0.00221 & 2.6 & \textbf{5}  & 0.01953 & 0.0212 \\
1.4  & 0  & 0.00195 & 0.00221 & 2.7 & 0  & 0.00195 & 0.00221 \\
1.5 & 0  & 0.00195 & 0.00221 & 2.8 & 0  & 0.00195 & 0.00221 \\
1.6 & 0  & 0.00195 & 0.00221 & 2.9 & 0  & 0.00195 & 0.00221 \\
1.7 & 0  & 0.00195 & 0.00221 & 3.0 & \textbf{9}  & \textbf{0.06445} & \textbf{0.0670} \\
\hline
\end{tabular}
\caption{Results of the Wilcoxon signed-rank test for point-wise comparison of equi-ensemble and weighted ensemble VQE, listing the statistics' values, raw $p$-values, and the $p$-values corrected via the Benjamini-Hochberg approach.}
\label{tab:pointwise}
\end{table}

\end{document}